\documentclass[a4paper,11pt]{article}
\usepackage{pos}
\usepackage{physics}

\usepackage[normalem]{ulem}  
\usepackage{color} 

\title{Threshold cusp structures in multi-channel scattering}

\author*[a]{Katsuyoshi Sone}
\author[a]{Tetsuo Hyodo}

\affiliation[a]{Department of Physics, Tokyo Metropolitan University,\\
  Hachioji 192-0397, Japan}

\emailAdd{sone-katsuyoshi@ed.tmu.ac.jp}
\emailAdd{hyodo@tmu.ac.jp}

\abstract{
We study the behavior of the cusp structures focusing on the isospin symmetry breaking effects. The properties of the exotic hadrons are reflected in the shape of the cusp structures. In realistic hadron scatterings, the threshold energies of the isospin partners appear within a small energy region. For a detailed analysis of such systems, it is essential to study the behavior of the cusp structures arising at two closely spaced thresholds. In this work, we introduce a convenient formulation of the scattering amplitude and demonstrate that the cusp structures at two nearby thresholds are related through the isospin symmetry.
}

\FullConference{The 21st International Conference on Hadron Spectroscopy and Structure (HADRON2025)\\
 27 - 31 March, 2025\\
Osaka University, Japan\\}

\begin{document}
\maketitle

\section{Introduction}

Many exotic hadrons appear as resonances near the threshold, and their properties are reflected in the shape of the scattering cross section~\cite{Esposito:2021vhu, LHCb:2021auc, ParticleDataGroup:2024cfk}. In particular, the threshold cusp structures encode valuable information about low-energy scattering dynamics and the nature of nearby resonance states~\cite{Baru:2004xg, Guo:2019twa, Dong:2020hxe, Sone:2024nfj, Zhang:2024qkg}.

In certain two-body hadronic systems, two thresholds may be located close to each other due to the isospin symmetry. In such cases, the two cusp structures appear in a narrow energy region. Here, we study the behavior of these cusp structures under the isospin-broken and isospin-symmetric cases, taking the $\Lambda N$-$\Sigma N$ coupled-channel system as a representative example.

\section{Formulation}\label{sec: form}

Before considering the effects of the isospin breaking, we first discuss the general form of the multi-channel scattering amplitude and analyze the behavior of the threshold cusp structures. The scattering channels are labeled as channels $1, 2, \cdots, N$, in ascending order of their threshold energies. The relative momenta in channels $1, 2, \cdots, N$ are denoted by $p_1(E), p_2(E), \cdots, p_N(E)$, respectively. The $s$-wave scattering amplitude is expressed as
\begin{align}
    f(E) = \left[\hat{K}^{-1} - i \hat{p}(E)\right]^{-1},
    \label{eq: f^N}
\end{align}
where $\hat{K}$ is an $N \times N$ real symmetric matrix known as the $K$-matrix~\cite{Badalian:1981xj}, and $\hat{p}(E)$ is a diagonal matrix with $(i, i)$ components given by $p_i\ (i=1,2,\cdots,N)$. 

We focus on the near-threshold energy region of channel $N$ and set the energy origin $(E=0)$ at this threshold. Because only the linear terms in $p_N(E)$ contribute to the cusp shape near the threshold of channel $N$, we neglect higher-order terms proportional to $p_N^2 \propto E$. Accordingly, $K$-matrix $\hat{K}$ and $p_i\ (i< N)$ in Eq.~\eqref{eq: f^N} are taken as constants. To analyze the cusp behavior, we represent the $(1,1)$ component of the scattering amplitude near the threshold of channel $N$ as
\begin{align}
    f_{11}(E) = f_{11}^{0} \frac{1 + i b_{11}^{(N)} p_N(E)}{1 + i a_N p_N(E)},
    \label{eq: amp of 11 near i}
\end{align}
which provides a convenient form for describing the cusp structure. Here, $a_N$ represents the scattering length in channel $N$, and $b_{11}^{(N)}$ is a complex constant with $\Im[b_{11}^{(N)}] \leq  0$, as required by unitarity. Both $a_N$ and $b_{11}^{(N)}$ are determined by the elements of $K$-matrix and the constant momenta $p_i\ (i<N)$. The factor $f_{11}^{0}$ denotes the amplitude at the threshold of channel $N$ $(E = 0)$.

\begin{figure}[tbp]
    \centering
    \includegraphics[width = 12cm, clip]{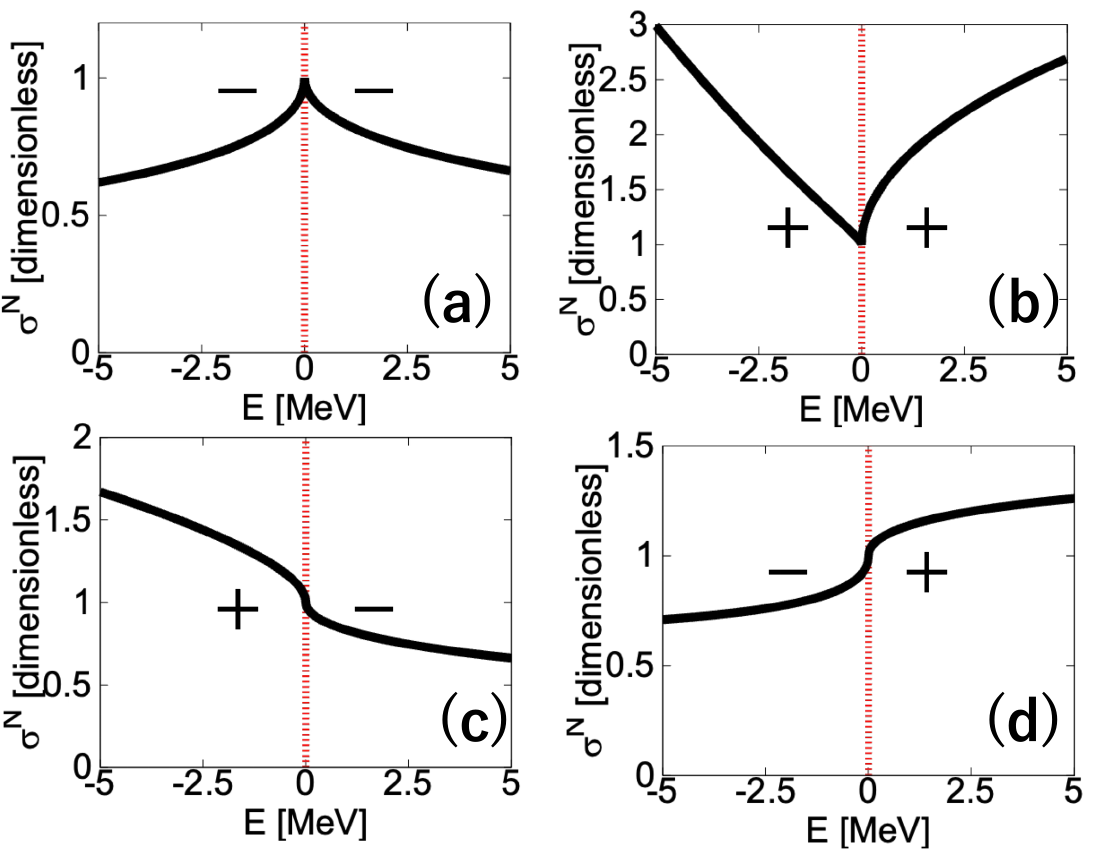}
    \caption{
    The normalized cross sections $\sigma_{11}^{\rm N}(E)=|f_{11}(E)|^2/|f^0_{11}|^2$ as functions of the energy $E$. Four panels correspond to $\qty{{\rm sgn}(\Im[a_N-b^{(N)}_{11}]),{\rm sgn}(\Re[a_N-b^{(N)}_{11}])}=\qty{-,-}$ (a), $\qty{+,+}$ (b), $\qty{+,-}$ (c) and $\qty{-,+}$ (d).
        }
    \label{fig1}
\end{figure}

Using Eq.~\eqref{eq: amp of 11 near i}, the $s$-wave cross section near the threshold can be expanded in terms of $p_N(E)$:
\begin{equation}
\begin{aligned}
    \sigma_{11}^s(E) &= 4\pi|f_{11}^{0}|^2 \left(1 + 2 \Im[a_N - b_{11}^{(N)}] p_N(E) + \mathcal{O}(p_N^2(E))\right), \quad (E > 0\ \text{above the threshold}), \\
    \sigma_{11}^s(E) &= 4\pi|f_{11}^{0}|^2 \left(1 + 2 \Re[a_N - b_{11}^{(N)}] \kappa_N(E) + \mathcal{O}(\kappa_N^2(E))\right), \quad (E < 0\ \text{below the threshold}),
    \label{eq: expansion of cses}
\end{aligned}
\end{equation}
with $\kappa_N(E) = -i p_N(E)$ which is real because $p_N(E)$ becomes pure imaginary for $E<0$. Note that the behavior of the cross section must be treated above and below the threshold separately. When we use the energy $E$ as a variable, the slopes of the cross section diverge at the threshold. Therefore, the shape of the cusp is determined only by the signs of $\Re[a_N - b_{11}^{(N)}]$ and $\Im[a_N - b_{11}^{(N)}]$. Depending on their signs, four distinct types of cusp structures may arise, as shown in Fig.~\ref{fig1}.

\section{$\Lambda N$-$\Sigma N$ scattering}
\label{eq: LN scattering}

As a representative example, we consider the $\Lambda N$-$\Sigma N$ scattering system with the total charge $Q=+1$. The scattering channels $\Lambda p$, $\Sigma^+ n$, and $\Sigma^0 p$ are labeled as channels 1, 2, and 3, respectively. Due to the isospin symmetry, the threshold energies of channels 2($\Sigma^{+}n$) and 3($\Sigma^{0}p$) are very close to each other, making this system particularly suitable for studying the isospin-breaking effects on the cusp structures. We define the energy difference between the thresholds of 2 and 3 as $\Delta$, which is about 2 MeV.

To analyze the $\Sigma N$ cusp behavior, we examine the $(1,1)$ component of the scattering amplitude. For $\Delta \neq 0$, two distinct cusps emerge at the separate thresholds of channels 2 and 3. We also consider the isospin symmetric case by taking the limit $\Delta \to 0$. In this case, the thresholds of channels 2($\Sigma^{+}n$) and 3($\Sigma^{0}p$) coincide, and the corresponding channel is denoted as I\hspace{-1.2pt}I($\Sigma N$). Thus, only a single cusp appears at the I\hspace{-1.2pt}I threshold.

In our analysis, the $K$-matrix for $\Lambda N$-$\Sigma N$ scattering is assumed to respect the isospin symmetry, which reduces the number of independent parameters. Namely, the $K$-matrix for $\Lambda N$-$\Sigma N$ scattering is described by four parameters:
\begin{align}
    K
    &=
    \begin{pmatrix}
        C_1 & \sqrt{2}C_4 & C_4 \\
        \sqrt{2}C_4 & C_2 & \sqrt{2}(C_2 - C_3) \\
        C_4 & \sqrt{2}(C_2 - C_3) & C_3
    \end{pmatrix}.
    \label{eq: Kmatrix for lambdaN}
\end{align}
The difference of the momenta $p_2(E)$ and $p_3(E)$, which arises due to $\Delta$, introduces the isospin-breaking effects into the scattering amplitude $f_{11}(E)$.

First, we consider the isospin-broken case. According to Eq.~\eqref{eq: expansion of cses}, the shapes of the cusps at the thresholds of channels 2 and 3 are determined by the signs of the real and imaginary parts of $a_2 - b^{(2)}_{11}$ and $a_3 - b^{(3)}_{11}$, respectively. Using the $K$-matrix components from Eq.~\eqref{eq: Kmatrix for lambdaN}, these quantities can be expanded in terms of $\kappa_3(\Delta)$ and $p_2(\Delta)$:
\begin{align}
    a_2 - b^{(2)}_{11} &= 2R + X \kappa_3(\Delta)+ \mathcal{O}(\kappa_3^2(\Delta))\quad (\text{threshold of 2}) ,  \label{eq: slope of th 2} \\
    a_3 - b^{(3)}_{11} &= R + W p_2(\Delta)+ \mathcal{O}(p_2^2(\Delta))\quad (\text{threshold of 3}), \label{eq: slope of th 3} \\
    R &\equiv -\frac{C_4^2}{(1 - i C_1 p_1) C_1},
\end{align}
where $X$ and $W$ are the constants determined by $C_{i}$ and $p_{1}$. Because $\kappa_3(\Delta)$ and $p_2(\Delta)$ vanish, only the first terms remain in the limit $\Delta \to 0$. The leading terms in Eqs.~\eqref{eq: slope of th 2} and \eqref{eq: slope of th 3} are identical except for a factor 2, which originates from Clebsch-Gordan coefficients. Therefore, if $2|R/X| \gg \kappa_3(\Delta)$ and $|R/W| \gg p_2(\Delta)$, the types of the cusps (see Fig.~\ref{fig1}) at the thresholds of 2 and 3 are expected to be identical.

In the isospin-symmetric limit, i.e., near the threshold of channel I\hspace{-1.2pt}I with $\Delta \to 0$, the slope of the cross section is given by
\begin{align}
    a_{{\rm I\hspace{-1.2pt}I}} - b_{11}^{({\rm I\hspace{-1.2pt}I})} = 3R.
    \label{eq: slopes of isospin limit}
\end{align}
Equation~\eqref{eq: slopes of isospin limit} shows that the slope in the isospin-symmetric case is the sum of $a_2 - b^{(2)}_{11}$ and $a_3 - b^{(3)}_{11}$ in the $\Delta \to 0$ limit.

\section{Numerical results}\label{sec: num}
 
In this section, we study the behavior of the threshold cusp structures in the $\Lambda N$-$\Sigma N$ scattering for both the isospin-symmetric ($\Delta\to 0$) and the isospin-broken ($\Delta \neq 0$) cases numerically. We use the averaged energy of $\Sigma^{+}n$ and $\Sigma^{0}p$ as the threshold energy of channel I\hspace{-1.2pt}I($\Sigma N$) for $\Delta\to 0$. The $\Lambda p$ elastic scattering amplitude $f_{11}(E)$ is calculated using Eq.~\eqref{eq: f^N} together with the $K$-matrix defined in Eq.~\eqref{eq: Kmatrix for lambdaN}. 

In our numerical analysis, we define a dimensionless cross section $\sigma_{11}^{\rm N}(E)$ as
\begin{align}
    \sigma_{11}^{\rm N}(E) 
    &= \frac{|f_{11}(E)|^2}{|f_{11}(E = 0;\, \Delta \to 0)|^2}, 
    \label{eq: csec norm}
\end{align}
where $f_{11}(E; \Delta \to 0)$ is the scattering amplitude in the isospin-symmetric limit. Accordingly, $\sigma_{11}^{\rm N}(E;\Delta \to 0)$ is normalized at the threshold of channel I\hspace{-1.2pt}I.

In order to determine the four parameters in Eq.~\eqref{eq: Kmatrix for lambdaN}, we need four conditions. First, we fix the scattering length $a_{{\rm I\hspace{-1.2pt}I}}$ in the $\Delta \to 0$ limit. The real and imaginary parts of $a_{{\rm I\hspace{-1.2pt}I}}$ impose two constraints on the parameters in Eq.~\eqref{eq: Kmatrix for lambdaN}. It is important to note that, within the first-order expansion in the momentum, the position of the pole is determined only by the scattering length in the $\Delta\to0$ limit.

In addition to this constraint, we impose two conditions $C_1 C_3 - C_4^2 = 0$ and $C_2 - 2 C_3 = 0$, which lead to $b_{11}^{(\rm I\hspace{-1.2pt}I)} = 0$. As a result, the slopes of the cross sections in Eq.~\eqref{eq: slopes of isospin limit} depend only on the scattering length $a_{\rm I\hspace{-1.2pt}I}$. These four conditions uniquely determine the four independent parameters in Eq.~\eqref{eq: Kmatrix for lambdaN}. We also use this parameter set in the isospin broken case ($\Delta\neq0$). The hadron masses used in the calculation are taken from Ref.~\cite{ParticleDataGroup:2024cfk}.

\begin{figure}[tbp]
    \centering
    \includegraphics[width = 8cm, clip]{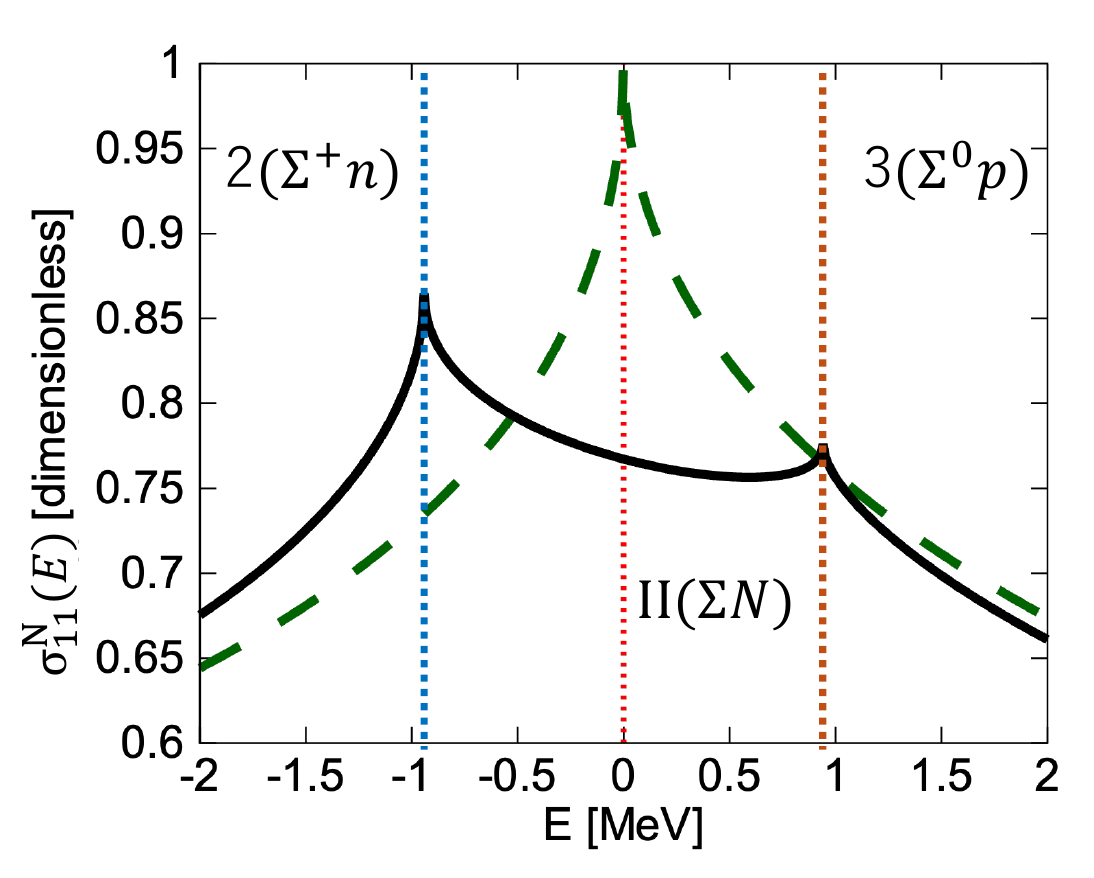}
    \caption{
        The (1,1) component of the cross section $\sigma^{\rm N}_{11}(E)$.  The solid and dashed lines correspond to $\Delta\neq0$ and $\Delta\to0$ cases, respectively. The dotted vertical lines represent the thresholds of channel 2 ($\Sigma^+ n$), I\hspace{-1.2pt}I ($\Sigma N$) and 3 ($\Sigma^0 p$).
        }
    \label{fig2}
\end{figure}

As an example, we use the scattering length of the typical hadronic scale for $\Delta\to0$;
\begin{align}
    a_{{\rm I\hspace{-1.2pt}I}} &= -1.0 - i\,0.8\quad {\rm fm}.
    \label{eq: a2s num}
\end{align}
The corresponding values of the scattering lengths $a_2$ and $a_3$ with $\Delta\neq 0$ are obtained as
\begin{align}
    a_2 &= -0.65-i0.46 \quad \rm fm, \label{eq: cp a2} \\
    a_3 &= -0.26-i0.27\quad \rm fm. \label{eq: cp a3}
\end{align}
For $a_{{\rm I\hspace{-1.2pt}I}}$ in Eq.~\eqref{eq: a2s num}, $f_{11}(E; \Delta \to 0)$ has a quasivirtual pole~\cite{Nishibuchi:2023acl} below the I\hspace{-1.2pt}I threshold. The resulting cross sections for both the isospin-symmetric and isospin-broken cases are shown in Fig.~\ref{fig2}. The dashed line represents $\sigma_{11}^{\rm N}(E)$ in the isospin-symmetric case ($\Delta \to 0$) and the solid line corresponds to the case with finite $\Delta$. 

From the solid line in Fig.~\ref{fig2}, it is observed that the cross section $\sigma_{11}^{\rm N}(E)$ for $\Delta \neq 0$ exhibits cusp structures at the thresholds of channels 2 and 3 which are  the same type of cusp with Fig.~\ref{fig1}(a), consistent with $\Im[a_N-b^{(N)}_{11}]<0$ and $\Re[a_N-b^{(N)}_{11}]<0$. The isospin-symmetric cross section $\sigma_{11}^{\rm N}(E)$ with $\Delta\to 0$, shown as the dashed line in Fig.~\ref{fig2}, displays the same type of cusp behavior as the solid line. This result indicates that the isospin-breaking effects are small for the case of $a_{{\rm I\hspace{-1.2pt}I}} = -1.0 - 0.8\ {\rm fm}$, as discussed in Sec.~\ref{eq: LN scattering}. Indeed, the sum of the scattering lengths $a_2 + a_3 = -0.91 - i\,0.73\ {\rm fm}$ is approximately equal to $a_{{\rm I\hspace{-1.2pt}I}}$, further supporting our conclusion.

\section{Summary}

In this contribution, we study the effects of isospin breaking on the threshold cusp structures. In Sec.~\ref{sec: form}, we introduce a convenient representation of the scattering amplitude $f_{11}(E)$ as given in Eq.~\eqref{eq: amp of 11 near i}, which allows for a transparent analysis of the slope of the cross section near the threshold. Using this representation, in Sec.~\ref{eq: LN scattering}, we examine the behavior of the $\Sigma N$ cusp structures in the $\Lambda p$ elastic scattering amplitude under both the isospin-symmetric and isospin-broken cases. As a result, we find that the types of the cusp structures at the thresholds of $\Sigma^+ n$ and $\Sigma^0p$ with $\Delta\neq0$ become identical to the single cusp with $\Delta\to0$, suggesting that the isospin breaking effect is small. We demonstrate numerically this phenomena in Sec.~\ref{sec: num}. 

\begin{acknowledgments}
This work has been supported in part by the Grants-in-Aid for Scientific Research from JSPS (Grants
No.~JP23H05439, 
No. JP22K03637, and 
 by the RCNP Collaboration Research network (COREnet) 048 "Revealing the nature of exotic hadrons in Belle (II) by collaboration of experimentalists and theorists", 
and by JST SPRING, 
Grant Number JPMJSP2156

\end{acknowledgments}


\end{document}